\documentclass[conference]{IEEEtran}
\usepackage{graphicx}
\usepackage{subcaption}
\usepackage{amsmath,amssymb,epsfig,xcolor} % easier math formulae, align, subequation
\usepackage{tabularx,multirow,array}
\usepackage{times}
\usepackage{booktabs}
\usepackage{multirow}
\usepackage{soul}
\usepackage{enumerate}
\usepackage[countmax]{subfloat}
\usepackage{threeparttable}
\usepackage{cite}
\usepackage{url}
\usepackage{amsthm}
\usepackage[utf8x]{inputenc}
\makeatletter
\def\endthebibliography{%
  \def\@noitemerr{\@latex@warning{Empty `thebibliography' environment}}%
  \endlist
}
\makeatother

\newcommand{\code}[1]{\texttt{#1}}

\begin{document}

\title{Engineering Edge-Cloud Offloading of Big Data for Channel Modelling in THz-range Communications}
\author{
    
    \IEEEauthorblockN{Zied Ennaceur, Anna Engelmann and Admela Jukan}
    \IEEEauthorblockA{Technische Universit\"at Braunschweig, Germany
    \\\{z.ennaceur, a.engelmann, a.jukan\}@tu-bs.de}
}
\providecommand{\keywords}[1]{\textbf{\textit{Index terms---}} #1}
\maketitle

\begin{abstract}
Channel estimation in mmWave and THz-range wireless communications (producing Gb/Tb-range of data) is critical to configuring system parameters related to transmission signal quality, and yet it remains a daunting challenge both in software and hardware. Current methods of channel estimations, be it modeling- or data-based (machine learning (ML)), - use and create big data. This in turn requires a large amount of computational resources, \code{read} operations to prove if there is some predefined channel configurations, e.g., QoS requirements, in the database, as well as \code{write} operations to store the new combinations of QoS parameters in the database. Especially the ML-based approach requires high computational and storage resources, low latency and a higher hardware flexibility. In this paper, we engineer and study the offloading of the above operations to edge and cloud computing systems to understand the suitability of edge and cloud computing to provide rapid response with channel and link configuration parameters on the example of THz channel modeling. We evaluate the performance of the engineered system when the computational and storage resources are orchestrated based on: 1) monolithic architecture, 2) microservices architectures, both in edge-cloud based approach. For microservices approach, we engineer both Docker Swarm and Kubernetes systems. The measurements show a great promise of edge computing and microservices that can quickly respond to properly configure parameters and improve transmission distance and signal quality with ultra-high speed wireless communications.
\end{abstract}

\section{Introduction}
%The terahertz (THz) frequency ($0.3-10$ THz) can provide high throughput to satisfy the exponentially growing data volumes in wireless networks\cite{petrov2016terahertz}. Since THz band is quite sensitive to molecular absorption in the atmosphere, resulting in low signal-to-noise ratio (SNR), there is a need for the channel estimation to properly configure THz sender and receiver to improve the signal quality. 

The channel estimation in mmw and THz-range of frequencies is a complex problem due to channel dependency on several physical factors. With the increasing proliferation of ML, the THz channel estimation especially can be performed by learning the patterns from previous experiments and periodical channel probing \cite{ye2017power}. To satisfy users' special requirements on quality and configurations of THz channel, a large amount of THz system parameters and computationally complex ML algorithms need to be controlled, managed and monitored. The complexity of control, management and monitoring in such a system also increases with the number of concurrent data used for channel estimation, resulting in a huge number of \code{read}, \code{write} and other computational operations. Currently, THz channel modeling produces Tbytes of Data. Further on the horizon, the control plane of THz networked systems are likely to require much more computational power and memory as compared to the local control computing units collocated with antennas \cite{shokri2015millimeter}. To address these challenges, the design of novel computation systems for control and management concepts in high-speed wireless communications is indispensable to provide real-time response and predictable performance of the system parameters, such that the channel quality is guaranteed. %and fast configuration of the THz system.%, while the main question is where to place and how to organize the computational resources and storage to be able to meet the requirements of the low response time, i.e., latency.

Offloading big data for channel estimation to edge and cloud computing systems appears as a straightforward solution idea.  At the same time, however, data stored and processed in the cloud can suffer from performance issues including delays due to its centralized and remote locations, which in turn can result in slower and non-real time response to THz system’s needs to timely configure antennas and transmission parameters. On the other hand, edge computing is envisioned to provide compute, storage and network services as close as possible to the transmission systems with a comparably lower latency overall \cite{8089336}. Since the edge-cloud solutions however require the orchestration middleware, one needs to consider additional delays that can impact response times. This in turn also requires innovation in  suitable orchestration mechanisms \cite{zhang2019hetero}. Today, all big-data approaches to channel estimation in communications use the traditional monolithic engine, which is easy to engineer as it is built as a single local unit, co-located with the system. However, its limited scaling capability results in a high blocking rate of requests as the system evolves. The container-based architectures, in contrast, can address the problem of scalability in case of channel modeling with a large number of concurrent requests \cite{bernstein2014containers}. Microservices can be deployed in both cloud and edge using one of the two most popular container management systems, i.e., Docker Swarm or Kubernetes \cite{rufino2017orchestration}. To engineer a computing and storage concept for channel estimation and other control plane functions, it is necessary to benchmark the performance of cloud and edge with different orchestration engines in terms of latency, throughput and scalability. 

We engineer an edge-cloud computing system for offloading of big data for channel modeling, and focus on a case study of THz-communications due to its abundant data and transmission speeds. Specially, we focus on analyzing and studying how the data-based (ML based) approach for dynamic channel estimation could solve the issues related to signal quality estimation of ultra-high speed communication systems. The channel estimation parameters are 
used for a timely channel configuration such that the system can dynamically configure and improve the signal-to-noise ratio (SNR). We first engineer a monolithic node to processing the requests from THz transmission system. We then focus on Docker Swarm and Kubernetes implementations. We experimentally benchmark the performance in terms of latency and throughput. The results indicate that edge node can provide a low latency using monolithic architecture. Cloud solution outperforms edge when the number of concurrent requests is limited. To the best of our knowledge, this paper is the first one to address the issues of engineering and offloading the THz channel modeling related big-data into the edge and cloud, a necessary step towards the feasibility of THz communications. 

The rest of the paper is organized as follows. Section II  describes the computing system architecture for the THz channel modeling. Section III shows the offloading to edge and cloud approach. Section IV shows the performance evaluation of the proposed system and Section V concludes the paper.

\vspace{-0.3 cm}
\begin{figure}[hbt!]
\centering
\includegraphics[width=0.9\columnwidth]{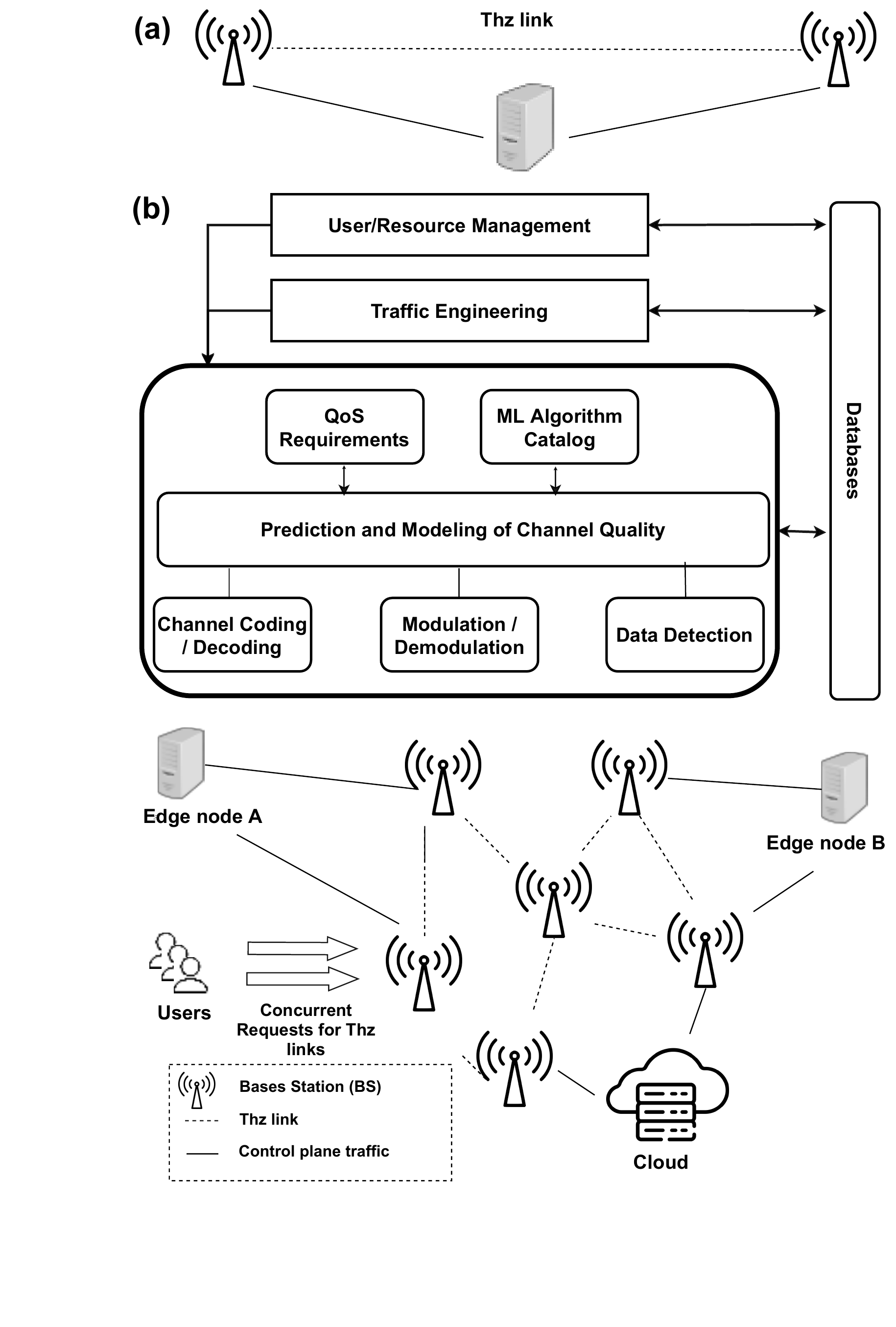}
\vspace{-1 cm}
 %   \vspace{-1.9cm}
    \captionof{figure}{(a) Today's THz link,  (b) Future THz mesh network.}
    \vspace{-0.5cm}
    \label{fig:net}
\end{figure}

%\section{Preliminaries}
\section{Channel Modeling in THz Communications}\label{UseCase}

Recent wireless communication systems aims at using high frequencies in order to provide high bitrates. To this end, new technologies have been considered to exploit high frequency ranges and deal with channel modeling and estimation problems, such as AI/ML \cite{bensalem2021effectiveness}, beamforming and multiple-input multiple-output (MIMO) \cite{saad2019vision} systems. In Fig. \ref{fig:net}a) and \ref{fig:net}b), we illustrate how THz channel modeling is performed today, and the vision. Today, channel modeling in short range (mostly indoors) THz link is performed based on measurements, which needs to be stored locally and processed by ML algorithms. The ML algorithms are used to predict bit error rates (BER), SNR \cite{bensalem2021effectiveness} and also type of modulation \cite{wang2020deep} that needs to be used for a certain bandwidth request.  In \cite{bensalem2021effectiveness} ML algorithms are used to estimate the channel state information which is used  for data detection in a communication system, where we measure the received signal at the receiver side after sending a set of pilot symbols. Afterwards, the data driven ML algorithms create the channel state matrix and computes the expected BER and SNR of the transmission system.

Figure \ref{fig:net}b) show how these systems are expected to evolve in the future. We envision ultra-high bitrate THz network systems,  where multiple THz base stations (BS) are interconnected building a  mesh THz network. A large number of users can simultaneously request a THz link by sending requests for the desired Quality of Service (QoS) parameters. The BSs receive users' requests and need to process them before THz link can be defined and established. That processing in the compute nodes includes control, management and evaluation steps enabling to decide which THz link is the most suitable in regard to, e.g., bandwidth, transmission distance and bit rate. The possible processing steps required are User and Resource Management, Traffic Engineering, QoS evaluation, collection and evaluation of measurements  as well as Channel Estimation. All these processing steps will trigger an amount of \code{read}, \code{write} and computational operations. For channel estimation which is based on computationally complex ML algorithms and channel measurements in real time, Tbytes of data that need to be processed. The data processing and ML algorithms can be outsourced to an edge node or remote cloud. With the help of virtual entities in the cloud, the cloud can  support the processing of Big data with on-demand elastic storage, network and computing capability. Edge computing can complement the cloud with a faster networking time.

%\subsection{Channel modeling in THz networking}
%\subsection{Networking Model}

%\hl{here this is the vision. Thz doesnot exist yet, incorporate channel modelin into the current Figure 1 is not see in the figure. This part should motivate the need to offload to cloud and edge. }

\subsection{Future THz Channel Control Model: An Example}
The proposed THz channel control model consists of  a control plane module implemented in an external computing node, and illustrated in Fig. \ref{fig:net}b). To provide control, management and channel estimation functions, the computing node can consist of four components: User and Resource Management, Traffic Engineering, Evaluation of Channel Quality using ML-Algorithms (ML\_Alg\_1...n) and Database. User Management is utilized to assign a new user to a specific THz channel. The request for user management will trigger some \code{read} operations from the database to define the number of users already allocated to THz channels between certain pair of BSs. The database will be also updated by \code{write}  operation when a new user is allocated to a certain channel. The Resource Management is the next processing step, which requires a \code{read} operation to access the information about the available and occupied bandwidth of existing THz channels. Since resource management request can contain information about released bandwidth, the database needs to be updated as well, resulting in several \code{write}  operations. The Traffic Engineering block will take care of all the communication between different nodes inside the THz network providing routing services. Thus, a set of suitable THz channels can be defined to meet a certain requirements on the end-to-end path. To this end, the traffic engineering request needs to contain information about sources and final destination and results in multiple \code{read} operations in computing node. To define a suitable THz channel, the key module is \emph{Prediction and modeling of channel quality}. This module utilizes information from all other control and management blocks as well as incoming information from BS such as QoS requirements and measurements of THz channel. The QoS requirements provide information about requested quality of THz channel, e.g., bit rate, bit error rate bounds or type of data that will be sent over THz channel. The channel measurements can include a current SNR values and environment state such as humidity, temperature, etc. Both, incoming measurements and QoS, are utilized for future channel quality evaluation and result in numerous \code{read} and \code{write}  operations. 

The Prediction and modeling of channel quality is the main control block to us, which requires to overcome an amount of training data and measurements and thus needs an amount of computational resources, \code{read} and \code{write}  operations. The \code{read} operations are required to prove, if there is some predefined THz channel configurations in the database, which relates to arrived QoS requirements and measured channel state. If there is no pre-configurations, the evaluation of channel quality uses machine learning (ML) techniques to accurately estimate the modulation format, bandwidth, channel coding, transmission distance to configure THz transceivers, i.e., to define suitable THz link. Based on information from database, other control blocks, the measurements provided and user's QoS requirements, the best ML algorithm with the best combination of relevant parameters will be assigned. The \code{write}  operations are required to store the new combinations of QoS parameters, channel state parameters and related THz channel configuration in the database. The response to the BS contain all required parameters for THz link configuration. 

It should be noted that this is one of numerous possible control plane architecture in future THz mesh network. In this paper, since the focus is on big data, we emphasize the channel prediction and modeling. A myriad of other scenarios can be envisioned, where also other modules require data and ML-based processing and are with no doubt equally relevant.

%
%\begin{figure*}[hbt!]
%\makebox[0pt][l]{%
%\begin{minipage}{1\textwidth}
%\centering
%\includegraphics[width=1\textwidth]{figures/ArchitectureMonoMicro.pdf}
%    \captionof{figure}{(a) Monolithic, (b) Microservice with Docker Swarm and (c) Microservice with Kubernetes}
%    \vspace{-0.5 cm}
%    \label{orchestr}
%     \end{minipage}}
%\end{figure*}

\section{Offloading to Edge and Cloud} \label{Offloading}
This section describes three concepts to engineer computing nodes for channel estimation, i.e., monolithic, microservices with Docker Swarm and with Kubernetes. Generally, each computing node requires two internal logical components to be engineered, i.e., Data Retrieval and Processing Module and Services Module. The Data Retrieval module allows to pre-process incoming data and make them available for the services used. The Services module contains applications to provide different services, such as \code{read} or \code{write} operations from or to database, respectively, estimation of the channel, evaluation of channel quality parameters, etc. The orchestration engine defines the operating principles of these modules and optimizes the usage of different services to provide scalability, high throughput and short response time.

\subsection{Monolithic Node Engineering}

\begin{figure*}[hbt!]
\makebox[0pt][l]{%
\begin{minipage}{1\textwidth}
\centering
\includegraphics[width=0.8\textwidth]{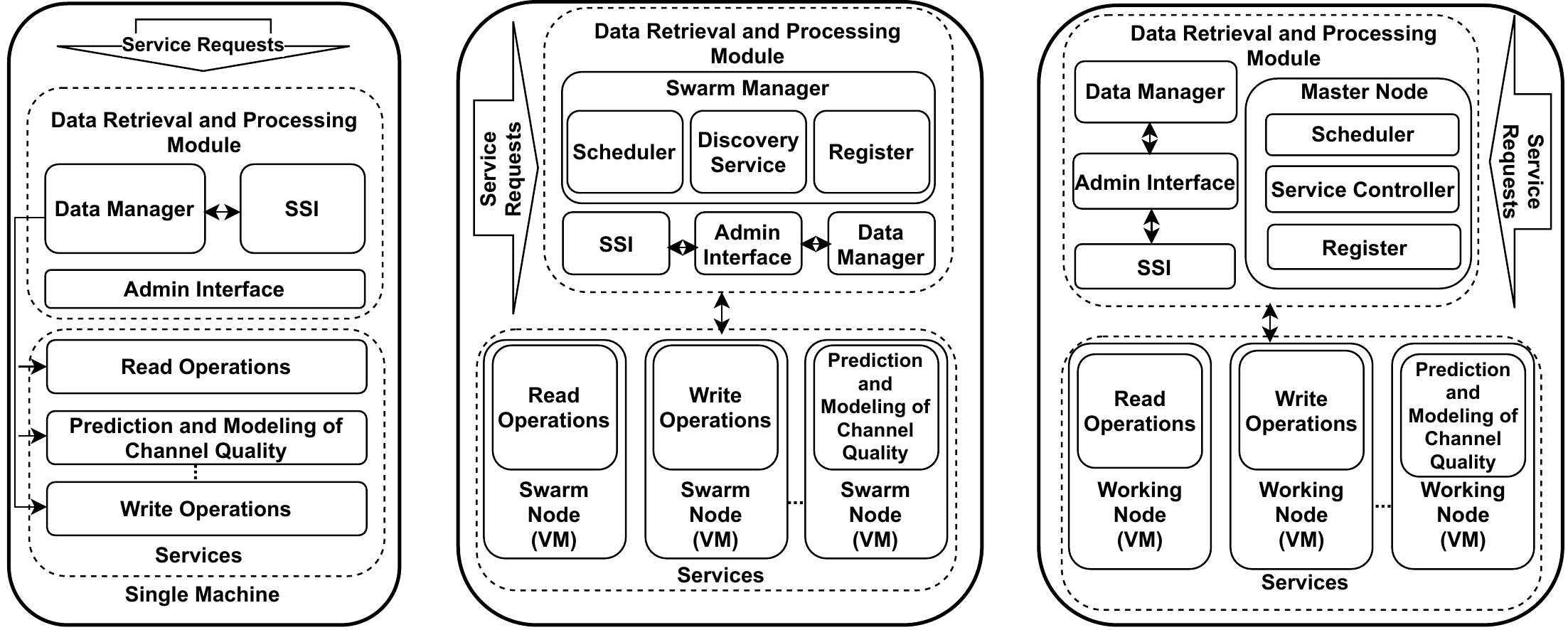}
    \captionof{figure}{(a) Monolithic, (b) Microservice with Docker Swarm and (c) Microservice with Kubernetes}
    \vspace{-0.5 cm}
    \label{orchestr}
     \end{minipage}}
\end{figure*}

The monolithic architecture is a traditional way to run services in computing nodes. In this implementation, the traditional architecture shown in Fig. \ref{orchestr}a) is simply offloaded to edge or cloud. Here, a single virtual machine (VM) builds a platform for data retrieval and processing, as well as for services. The Data Retrieval and Processing Block has a simple structure containing Data Manager and Service State Information (SSI) file. Data Manager is a part of the Model View Controller (MVC) middleware responsible for assignment of different data to the suitable service triggering several \code{read} and \code{write} operations, such as launching subsequently services for evaluation of channel quality. Additionally, Data Manager ensures the processing of requests during a predefined time to guarantee the rapid response. The SSI is used to define settings and preferences for building and running the services, such as the minimum amount of computational and storage resources allocated to each service, the number and type of services.
Generally in communication systems any incoming request needs multiple evaluation and processing steps. In typical systems we have a request to transmit a high bandwidth data over a suitable THz link, among many such links. This  requires a network management including resource management. 
Resource management allows an access to database and, thus, to information about currently available and allocated bandwidth, an other systems parametrs that neeed to be configured and utilized for THz channel estimation.
To configure the THz transmitter, a request is sent to the data manager, which collects, analyzes it and decides the type of data required for the services, i.e., first \code{read} from database and then \code{write} operation to update the database. Thus, first, the data manager will send an alert to notify the services component to activate the \code{read} service. The \code{read} service will connect to the database to \code{read} the values of configuration parameters and then send this value to the data manager. Data manager analyses the returned value and triggers further data retrieval and processing steps (if needed). When the channel can be estimated and configured based on available data, the data manager activates for instance the \code{write} operation and starts the channel estimation service; otherwise, a \code{read} retry starts after a certain time interval. After the channel is estimated, the data manager sends a response to the transmitter with all estimated configuration parameters for a THz link to be setup.
In general, a computing node can receive many requests at the same time,  which can lead to overload. A request for channel estimation on a certain link can also fail due to insufficient  computational and storage resources allocated. This is not only a problem for the computational node serving a certain transmission system, but can also create inconsistencies and blockages in high level management functions of the system. %For instance, while waiting for the computing node to respond, concurrent service requests remain underserved, and other, potentially valid links may remain unused for the duration of computation. 
Finally, it should be noted that monolithic implementations require redundancy or duplication of the entire code, to avoid services interruptions in case of failures. 

%Another reason for service interruption can be a missing service. When a new service is required to process an arrived request, the application manager will notify that new request and trigger a modification in the configuration file. After adding the new service the config file will alert the application manager to restart other services. That results in downtime of computing node, i.e., in additional delays of response and reduced throughput.

\subsection{Microservices}
In contrast to monolithic computing nodes, the microservice concept allows implementation of independent, small, scalable and flexible services allocated in different hosts and virtual machines (VM) referred to as \emph{nodes}. Each node owns an amount of resources adapted to the number of concurrent requests and the amount of requested computational and storage resources. To implement services on VMs providing a certain functionality, there is a need for so called containers which are a software components with a code, runtime, system libraries and settings required to run a certain service. Management and orchestration of the services across different VMs are the main challenges and require efficient orchestration tools such as Docker Swarm and Kubernetes.

\subsubsection{Docker Swarm}
Docker Swarm is an orchestration tool which allows to manage a group of virtual machines, i.e., Swarm nodes. Fig. \ref{orchestr}b) illustrates this architecture, which consists of Data Retrieval and Processing Block and Services Block. This block contains one primary and three secondary components. The primary component is Swarm Manager which includes Register, Scheduler \cite{scheduler} and Service Discovery. Scheduler analyses incoming requests and forward them to the requested service using the BinPacking strategy to decide in which node to run the requests. The Service Discovery monitors the services, their number and state as well as alerts administrator. The register buffers requests that can not to be processed immediately and need to wait for a service. The secondary components are Administration Interface, Data Manager and SSI. Administration interface allows an administrator an access to the Data Retrieval and Processing block. The Data manager collects and analyzes the received requests and decides the type of data required for the services. The SSI is a configuration file that contains information about the state of the services such as operating system, amount of swarm nodes, service availability, the allocated computational and storage resources, etc. The SSI contains some constant configurations of initial system state and dynamic part with configurations adapted to the system requirements and updated on-demand. The Services block is a set of VMs, i.e., swarm nodes, installed on different servers, whereby each VM allocates a certain service.

The request of a bandwidth in the Docker Swarm  is  forwarded to Swarm Manager, which  accesses SSI to obtain information about types and state of services, while Scheduler analyses the arrived request to determine the service, e.g., \code{read}. When the appropriate service exists and is ready to process requests, the Scheduler will activate that service, i.e., reading the value of allocated bandwidth from the database. The \code{read} service returns this value back to the same service, which triggers further processing steps, e.g., \code{write}  operation to reserve the bandwidth for a new THz link and evaluation of channel quality. When there is no available bandwidth, there is a need for a \code{read} retry after predefined time interval. After completion of processing, the Swarm Manager sends a response to the THz BS.

\subsubsection{Kubernetes}
Kubernetes is an open-source orchestration tool. Also here we implement Data Retrieval and Processing as well as Services. The Services block is a set of VMs so called Working nodes, allocated by different hosts. Each Working node uses Docker container to allocate a service, e.g., \code{read}, \code{write} , etc. In contrast to the Docker Swarm, Kubernetes deploys Master Node for Data Retrieval and Processing allocated in a separate VM. The main functions thereby are Service Controller, Data Manager, Scheduler \cite{schedulerkub} and Register.
Similar to docker swarm the Data manager will collects and analyzes the received requests and decides the type of data required for the services and send these informations to the Master Node.
The Scheduler receives all requests, evaluate them to predict required type of services and communicate with
register in case a service is not ready .
The Data Manager monitors a number of waiting requests and is able to create new Working nodes with required service using the Admin interface, as well as remove unused Working nodes to release computational and storage resources. The Scheduler forwards waiting requests for a certain services to the responsible Working node. SSI contains information such as type of services and container configurations, allocated resources, number of working nodes, etc. and is used by Data Manager.

As any request for a suitable THz link requires resource management, i.e., information about available and allocated bandwidth, the access to the database results in \code{read} and \code{write}  processing steps. Each incoming request is, first, sent to the Data Manager for pre-processing then to the Master node to allocated required service. Then, Scheduler forwards the request to the working node with a certain service, e.g., \code{read}. The appropriate working node receives the \code{read} request and activates the \code{read} service. After the \code{read} service is completed, the service sends the value of available bandwidth to the master node, which analyses the returned value and triggers further processing steps, e.g., \code{write}  service. Otherwise there is a need for a \code{read} retry.

When a large number of requests for the same service arrive and the service is overloaded, the Service Controller add an additional pods with a service of the same type as overloaded service or will use the autoscaling feature of the full cluster, without any service interruption.

%%%%%%%
\section{Implementation and Measurements}
\begin{figure}[t]
\centering
\includegraphics[width=0.8\columnwidth]{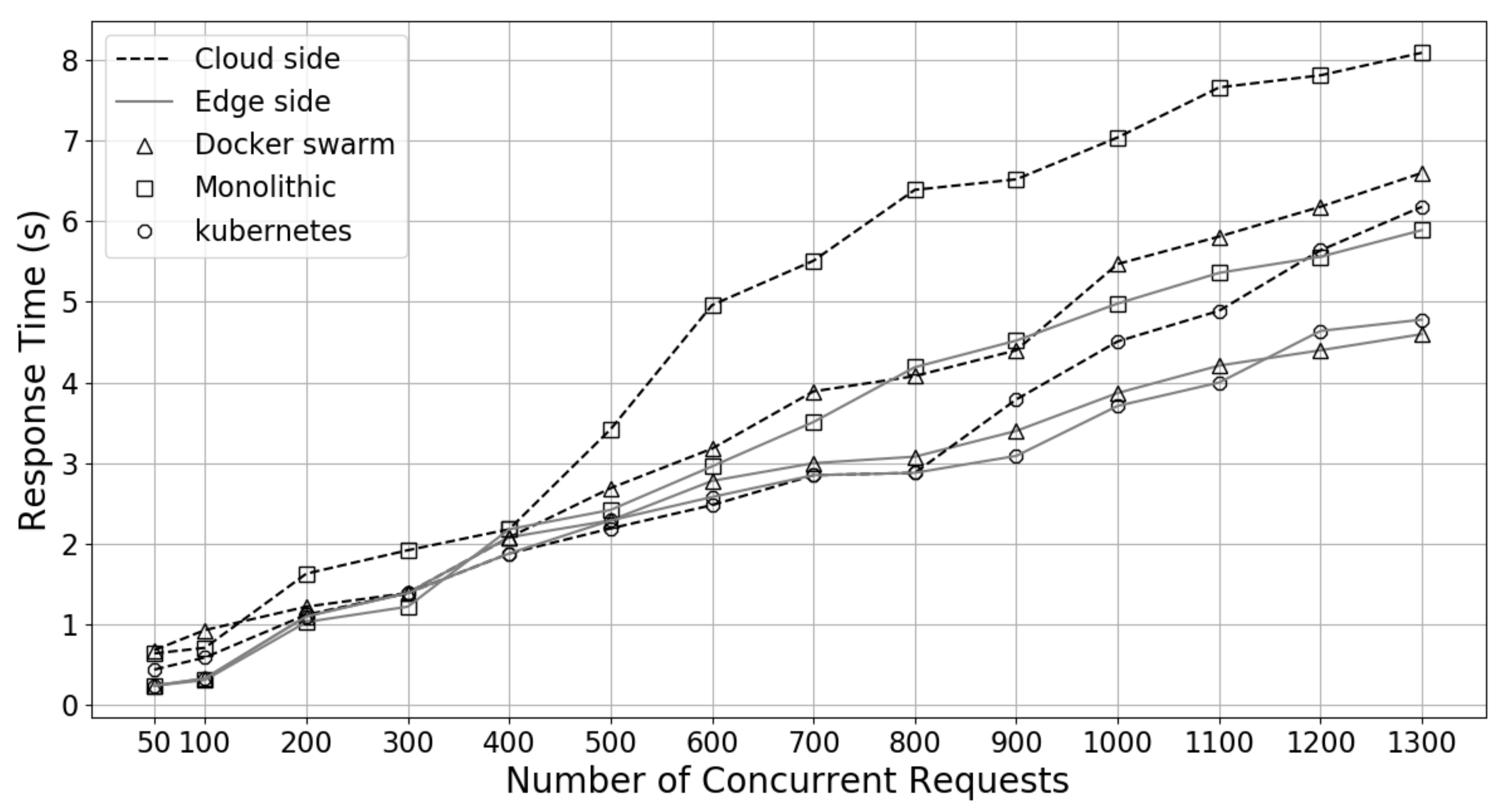}

  \caption{Response time for Read operation.}
  \vspace{-0.6cm}
  \label{TimeRead}
\end{figure}
\begin{figure}[t]
\centering
\includegraphics[width=0.8\columnwidth]{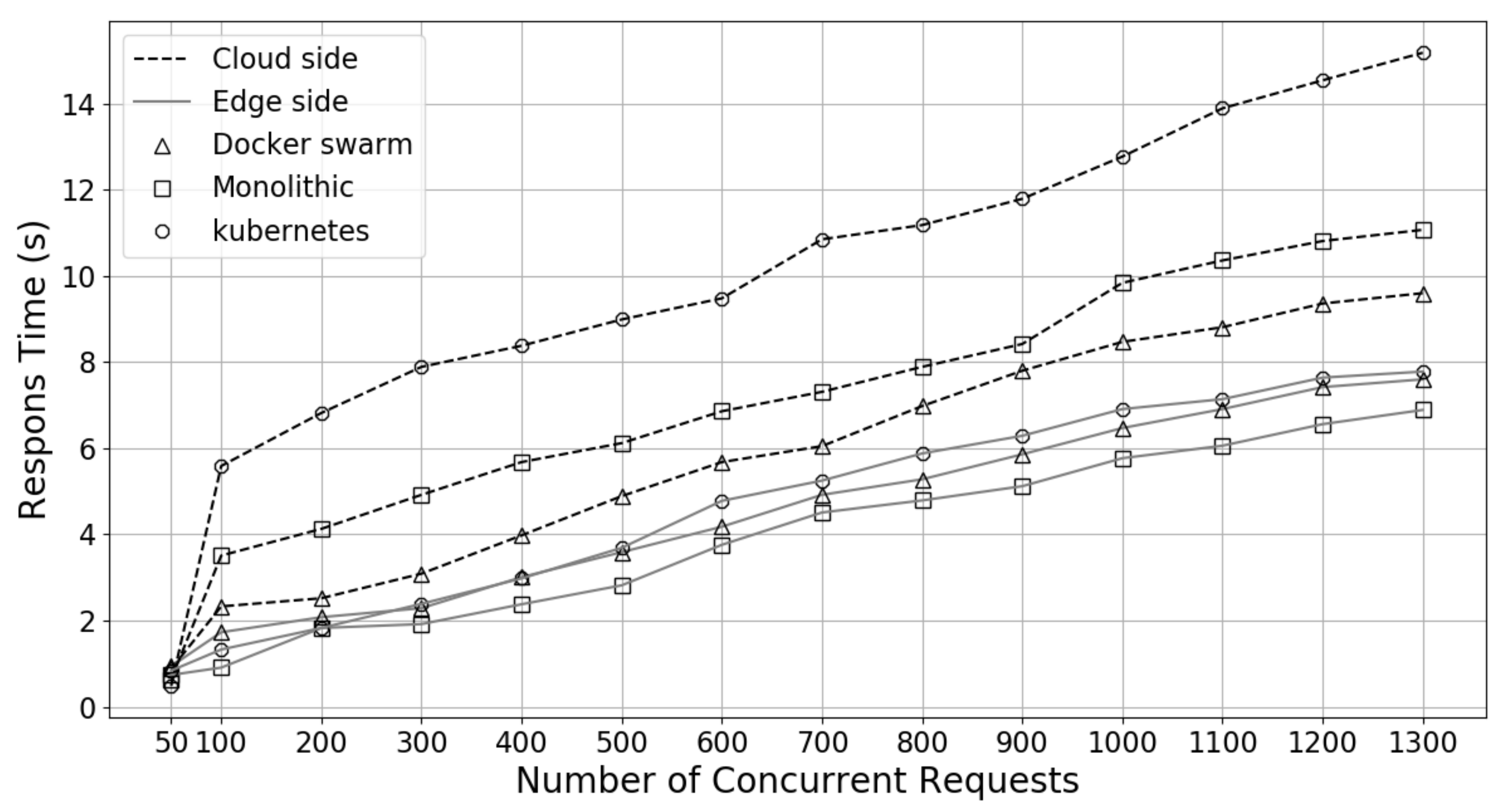}
  \caption{Response time for Write operation.}
  \vspace{-0.6cm}
  \label{TimeWrite}
\end{figure}

\begin{figure}[t]
\centering
\includegraphics[width=0.8\columnwidth]{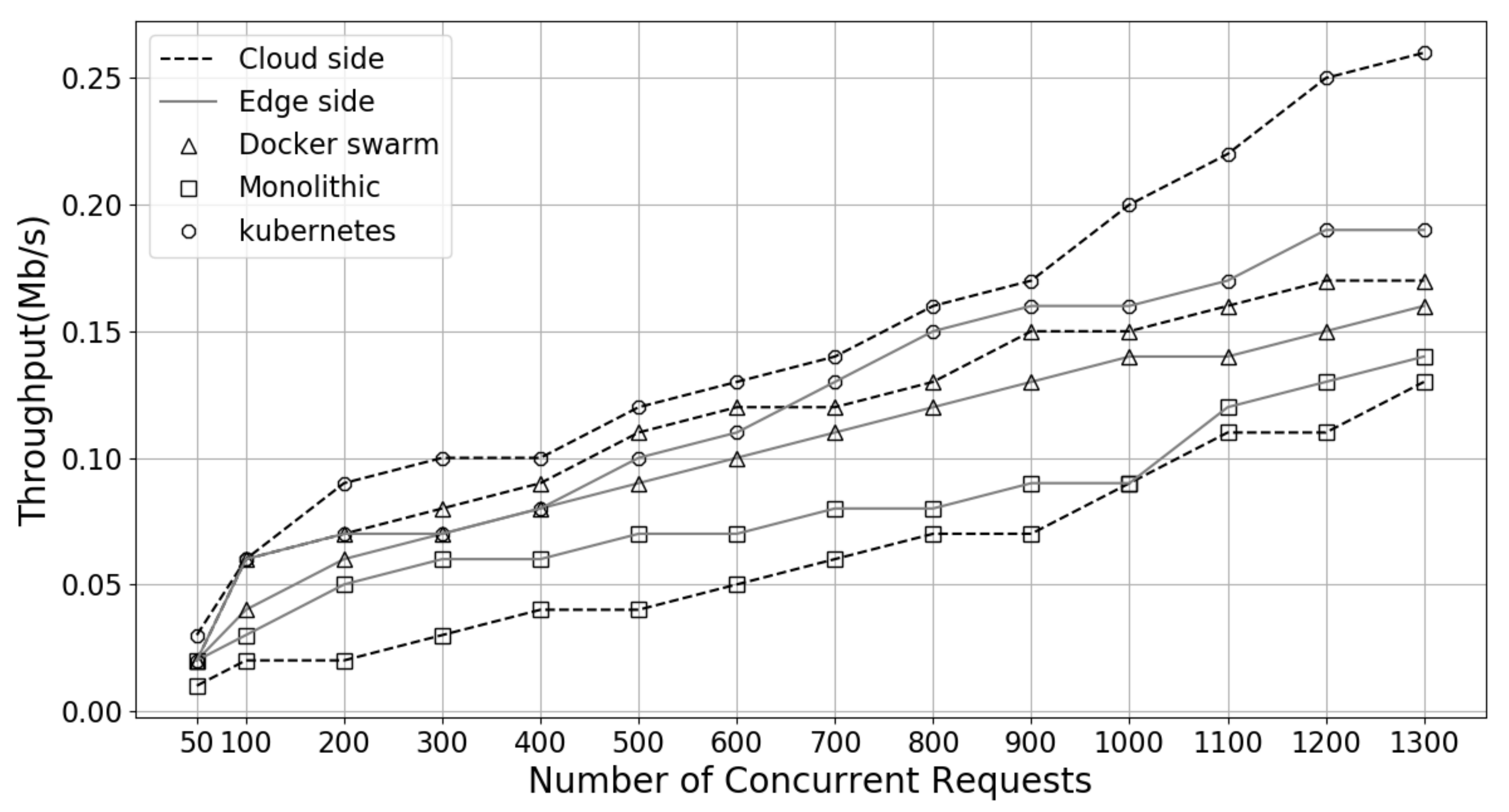}
  \caption{Throughput for Read operation.}
  \vspace{-0.6cm}
  \label{ThrRead}
\end{figure}
\begin{figure}[t]
\centering
\includegraphics[width=0.8\columnwidth]{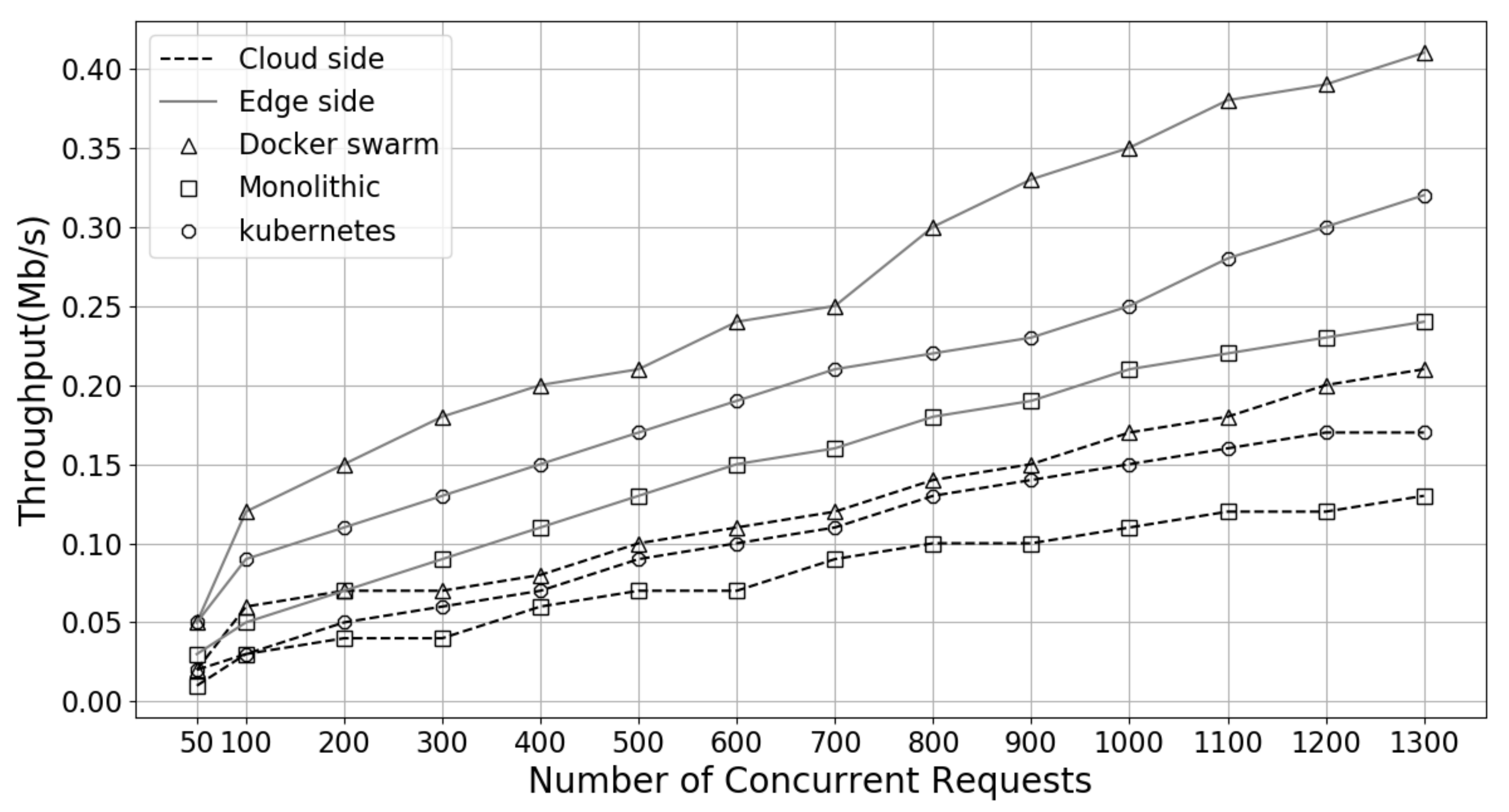}
  \caption{Throughput for Write operation.}
  \vspace{-0.6cm}
  \label{ThrWrite}
\end{figure}

We engineered an THz channel estimation offloading application using the three architectures i.e., monolithic, microservice with docker swarm and, kubernetes. Within we consider five main components: (i) admin interface service that contains a form to use by the users to insert the QoS parameters, (ii)
database service to store the channel measurement data, (iii) \code{read}, (iv) \code{write}  services, i.e.,  that \code{read} the data already stored in the database service and store it in another PostgreSql database. The last component is a NodeJs service to conduct the measurements towards performance results.

\par For the monolithic architecture we used the MVC pattern to offload our application. The MVC architectural pattern contains three main parts Model, View and Controller. Using this pattern the application is divided into three logical parts, the Model component and is connected to the database so anything we do with the data is done in the model component, the controller contains the data manager and the SSI of the data retrieval and processing module, the SSI is a JSON file that contains informations about the types of data allocated to each service and the view is our admin interface that generates user interface for the user and is created using the data collected by the Model component. All these components are built as a single and indivisible unit what make it simple to deploy but hard to update its components.
\par For microservices we deploy our application in both Swarm and Kubernetes clusters. The swarm cluster consists of 3 linux host machines which can communicate over a network together: One act as a manager and the two others act as workers. We created different services for different components already described above.
After initializing our clusters we deploy our services to the swarm in the form of a Compose file.
The swarm manager is the responsible of deploying the services and allocating the desired service to the dessired node using the scheduler. Each service is allocated in a specefic node using a single virtual machine. For kubernetes we used the same architecture as docker swarm and we created a cluster of three nodes, i.e., master and two workers. We created different Pods for different clustered services similar to docker swarm. Each service is allocated to a specific node (virtual machine). All components of the microservice architecture are independent from each other. That significantly increase robustness of control plane and service reliablity. For instance, a failure of write  service will not impact other services.

The testbed of our scenarios consists of two local desktop (both with i5-7600 CPU and 16GB of RAM ).
For the edge we have two nodes one is running as an edge node and the other one as a tester node. For the cloud we used GCP (Google Cloud Platform) as real cloud.
The edge node is running the three architectures: Monolithic, microservice with docker swarm and microservice with kubernetes.
Each microservice architecture running a cluster of three nodes one as a master and two as workers .
A total of 5 services are stack in each cluster as described before in \ref{Offloading}.C. For the monolithic architecture the services are deployed in one single machine.
The computer running as a tester node is running the siege client \cite{siege} as an open source regression test and benchmarking utility, and also stores the results of the measurements.
We evaluate these three architectures using the two machines and the cloud.
In both scenarios the tester node send HTTP POST requests in case of writing operations and HTTP GET requests in case of reading operations, to the cloud/Edge endpoint and uses the master node as a receiver for the microservice architectures, i.e., docker swarm and kubernetes, and the entire application URL for the monolithic application.

We performed the tests to measure the response time, the throughput and scalability. The response time is the time that the data took to reach the destination (cloud or edge) from the tester node. The throughput is the average number of bytes transferred every second from the server to all the simulated users. The scalability is estimated based on the response time results, whereby the tester node simulates from 50 to 1300 concurrent requests.
%by sending payload size of 5MB.
\par For response time results, we measure \code{read} and \code{write} operations. We then compare the results for all the three architectures previously described. Fig. 3 and Fig. 4 show the response time of the system for different number of users (i.e., processes),  respectively, for \code{read} and \code{write} operations when performing requests with two different computing nodes edge and cloud. The average latency increases exponentially with the increasing number of users in the system when we have \code{read} operations. For read operation, this behavior is more evident above 500 concurrent users when deploy monolithic in cloud and 900 users when deploy other architectures in edge as well in cloud. The reason behind can be explained by the early overload of the monolithic architecture and the variation of response time for every request. The response time, with 400 users or less, when using cloud, and with 600 users or less, when using edge, is similar in both cases independently from the orchestration tool.
The microservice architectures in edge shows better results compared to the monolithic in cloud and edge. This trend could be interpreted by the fact that microservice architectures are more flexible due to the independence of different services deployed.

In contrast, the results for the \code{write} operations show more differences between architectures studied. The data offloaded to cloud or using of kuberneets resukt in the largest reponse time.
However , the shortest response time can be reached by data offloading to edge with monolithic architecture.

\par Fig. 5 and Fig. 6 shows the throughput for read and write operations. The throughput results are showing similar results under 900 users for the microservices architectures. Here, it is important to notice that the system scales really well up to 700 requests. Above 700 requests show how the increment of number of requests exponentially increases the throughput to compute them in all three architectures.
Above the 900 users this trend became more obvious with kubernetes in the cloud this due to the compatatiblity of GCP and the Kubernetes architecture. In Fig. 6 we show throughput, but in this case for the writing operations described in Fig. 2. The differences between different architectures deployed are made more clear here. In all the tests performed with high load the microservices showed better results than monolithic.

\section{Conclusion} \label{sec:conclusion}
We engineered and studied the offloading of THz channel estimation problem to edge and cloud computing systems to understand the suitability of edge and cloud computing to provide rapid response with channel and link configuration parameters on the example of THz channel modeling. The measurements showed a great promise of edge computing and microservices that can quickly respond to properly configure parameters and improve transmission distance and signal quality with ultra-high speed wireless communications. We showed that edge based computing outperforms the cloud based implementation in terms of latency using the microservice architectures. In terms of throughput, the cloud with the microservice architectures outperforms the edge when the load is quite low. Future work needs to focus on computing needs of mesh THz network and novel control plane architectures for these critical systems in 6G communications.

%\section*{Acknowledgment}
%This work was partially supported by the DFG Project Nr. JU2757/12-1, "Meteracom: Metrology for parallel THz communication channels."

\bibliographystyle{IEEEtran}
\bibliography{IccBib}

\end{document}